\documentstyle[11pt,newpasp,twoside,epsf]{article}
\newcommand{\mhi}{M$_{\rm HI}$}

\newcommand{\mmol}{M$_{\rm mol}$}

\def\edcomment#1{\iffalse\marginpar{\raggedright\sl#1\/}\else\relax\fi}
\reversemarginpar

\begin{document}
\title{Star Formation in Tidal Dwarf Galaxies}
 \author{Ute Lisenfeld}
\affil{IRAM, Avenida Divina Pastora 7, N.C., 18012 Granada, Spain}
\author{Jonathan Braine, Olivier Vallejo}
\affil{Observatoire de Bordeaux, UMR 5804, CNRS/INSU, B.P. 89, F-33270 
Floirac, France}
\author{Pierre--Alain Duc}
\affil{CNRS and CEA/DSM/DAPNIA Service d'Astrophysicque, Saclay, 91191 
Gif sur Yvette Cedex, France}
\author{St\'ephane Leon}
\affil{Physikalisches Institut, Universit\"at zu K\"oln, Z\"ulpicher Str. 77,
50937 K\"oln, Germany
}
\author{Elias Brinks}
\affil{Departamento de Astronom\'\i a, Apdo. Postal 144,
Guanajuato, Gto. 36000, Mexico}
\author{Vassilis Charmandaris}
\affil{Cornell University, Astronomy Department,
Ithaca, NY 14853, USA}

\begin{abstract}
Tidal Dwarf Galaxies (TDGs) are objects presently 
forming from gas which has been expelled from their
parent galaxies during an interaction. We observed the CO emission
of  a sample of 11 TDGs, of which 8 were detected. 
The CO
is found at the peak of the HI observations and has the same
line velocity and width, indicating that the molecular gas is forming
in situ instead of being torn from the parent galaxies. The presence
of H$\alpha$ emission furthermore shows that stars are forming from this
molecular gas. In order to investigate star formation in TDGs further,
we compared their molecular gas content and 
star formation rate (SFR), traced by 
H$\alpha$, to those of spiral galaxies and classical dwarfs.
The major difference between TDGs and classical dwarfs is the 
lower metallicity of the latter. The star formation efficiency 
(SFR per molecular gas
mass) of TDGs lies in the range typical of spiral galaxies
indicating that star formation is proceeding in a normal fashion
from molecular gas.
%

\end{abstract}

\section{What are Tidal Dwarf Galaxies?}

TDGs are small galaxies which are currently in the
process of formation. They are forming 
from material ejected from the disks of spiral galaxies through  galactic 
collisions.
Their properties are very similar to classical dwarf irregulars
and blue compact dwarf galaxies, except for their metallicities which
are higher and lie in a narrow range of 12+log(O/H)$\approx 8.4 - 8.6$
(Duc et al. 2000). These metallicities
are typical of the material found in the outer
spiral disks of the interacting galaxies and which is
most easily lost to form TDGs.

TDGs are interesting objects because
they allow us to observe the process of galaxy formation, similar to
what occurred in the early universe, but in local objects at high
sensitivity and resolution. Their age can be constrained via numerical
simulations of the collision. 
Their discovery has raised the question how large a fraction
of the current dwarf galaxy population could have formed as
TDGs, and how these could be distinguished from the population
of what are believed to be field dwarf galaxies. A possible
way to distinguish between a field or tidal dwarf galaxy is
via their Dark Matter content.
According to simulations, TDGs should contain no dark matter (Barnes
\& Hernquist 1992), whereas most classical dwarfs have a high dark
matter fraction (see Hunter et al. 2000).


\section{CO in Tidal Dwarf Galaxies}

TDGs have a high SFR, their metallicities are not as
low as in many classical dwarfs and they possess high atomic hydrogen
column densities. Therefore, large amounts of molecular gas should be
present and, assuming that the conversion factor between the CO emission and
molecular gas mass is similar to the Galactic value,
it should be detectable.
Surprisingly, first attempts to detect CO in Arp 105 (Duc \& Mirabel 1994) 
and the Antennae (Smith \& Higdon 1994) failed.

Between June 1999 and September 2000 we observed a sample 
of 9 TDGs with the IRAM 30m telescope in CO(1--0) and CO(2--1). 
Two further TDGs, NGC 5291N and NGC 5291S,
were observed with the SEST telescope in the same transitions.  
The main motivations
for these observations were (i) to find out whether 
molecular gas can be detected in TDGs or whether their molecular
gas content is much lower than would be expected from simple
estimates, (ii) to study the formation of molecular gas from atomic gas,
(iii) and to to check whether star formation
is going on in the same way  as in other galaxies.
The details of the observations and  a broader discussion of the results
can be found in Braine et al. (2000, 2001). 

We searched for CO at the peaks of the HI emission and detected 8 of the
objects.
In Table 1 the observed TDGs, the measured 
molecular mass and further data are listed. 


\begin{table}[t]
\caption{Data for the sample of TDGs}
\begin{tabular}{lcccc}
\hline
Source/System  &  \mmol$^a$ & \mhi$^b$  & L$_{H\alpha}$
& metallicity \\ 
 & 10$^8$ M$_{\sun}$ & 10$^8$ M$_{\sun}$ & 
10$^{39}$ erg s$^{-1}$ & 12+log(O/H)\\
\hline
UGC957/NGC~520  & $\la$ 0.06 & 0.27  & -- & -- \\
NGC2782/NGC~2782& $\la$ 0.12 & 1.5  & -- & -- \\
Arp245N/Arp~245 & 0.64$^c$   & 2.7  & 7.4 & 8.6 \\
Arp105S/Arp~105 & 2.2        & 4.1 & 10 -- 20 & 8.4 \\
NGC4038W/Antennae & 0.03$^d$& 0.17$^d$  &1.7 & 8.4 \\
NGC4676N/NGC~4676 & 1.1        & 16   & 10 -- 20 &--  \\
NGC5291N/NGC~5291  & 1.9        & 24   & 9.3 & 8.4 \\
NGC5291S/NGC~5291  & 2.9        & 14  & 5.6 & 8.5 \\
IC1182/IC~1182  & $\la$ 1.0  & 23   & --  & 8.4 \\
NGC7319E/Stephan's Quintet & 4.5    & 8.7  & 13.6 &-- \\
NGC7252W/NGC~7252 & 0.36$^d$        & 0.78   & 1.0 & 8.6 \\
\hline
\end{tabular}
$^a $ The molecular gas mass
was calculated using a conversion factor of $N(H_2)/I_{CO} = 2\times 10^{20}$ 
cm$^{-2}$(K km s$^{-1})^{-1}$ and assuming a helium mass
fraction of 0.27. This yields 
$ M_{\rm mol} = 1.073 \times 10^4 (D/{\rm Mpc})^2 
(S_{CO}/{\rm Jy\, km\, s^{-1}}) M_{\sun} $

$^b$ Atomic (HI and He) gas mass  within the same area as CO.

$^c$ extended source, molecular and atomic gas masses in central beam only

$^d$ Derived from CO(2--1) line (CO(1--0) was not detected)
assuming a ratio of the brightness temperatures of CO(2--1)/CO(1--0) 
of 0.75 and
assuming that the CO emission fills the CO(2--1) beam. \mmol\, should
be viewed as an upper limit, 
because the CO(1--0) emission could be 
optically thin or even more concentrated.

\end{table}

\section{What is the origin of the molecular gas?}

At the scales sampled by the CO and HI observations, 1 -- 10 kpc, 
the CO is detected at the HI column density peak and shares the
same kinematics in terms of line widths and velocity.
This is very different to what is found in spiral galaxies,
where CO is only detected within the optical radius
whereas HI is  much more broadly distributed (e.g. NGC~891,
Garc\'\i a--Burillo 1992).
This suggests that the molecular gas that we are observing
in the TDGs is {\em not} torn from the parent galaxies 
but formed {\it in situ}. In the following we will give some more
arguments why we think this is so and present counterarguments why we
consider it unlikely that the molecular gas has been torn as such from
the parent galaxies, together with the atomic gas.

First of all, based on the standard dust model (Mathis,
Rumpl \& Nordsieck 1977), the
timescales to transform, say 20\% of HI into $\rm H_2$ is of order 
$t_{20\%} = 10^7/n_{\rm HI}$ yr. At HI volume densities of typically $>$ 1 
at cm$^{-3}$ this is
much shorter than the timescale of the interaction which is of order
$10^8$ yr. Secondly, if the molecular gas were torn from the outer disks
of the interacting galaxies we need to assume that there exist large
quantities of $\rm H_2$ beyond the optical radius of a typical spiral galaxy
which is hidden from view, either because: 
(i) the gas has a too low density, (ii) it is too cold or (iii) its
metallicity is too low. All three possibilities seem unlikely:




(i) If the molecular gas had very low density, CO could be dissociated.
However, CO becomes self-shielding at CO column densities of a few $10^{14}$
cm$^{-2}$, corresponding to typical H$_2$ column densities of
several 10$^{20}$ cm$^{-2}$. Absorption measurements towards stars indicate
that at  total (atomic and molecular) hydrogen column densities of 
order a few 10$^{20}$ cm$^{-2}$ the molecular hydrogen fraction is 
very small (Federman et al. 1979). Thus, if the gas is dense enough to 
form molecular clouds, it is also dense enough to have a normal CO content.


(ii) Liszt \& Lucas (1998 and references therein) observed 
Galactic CO in emission and in absorption towards quasars.
In virtually all cases where absorption was measured, they also
found CO emission. This shows that there is no  substantial 
population of molecular clouds in the Galaxy where the CO is present but too 
cold to be detected in emission.


(iii) Is it possible that TDGs are formed from molecular gas
of the very outer regions of spirals where low metallicity 
inhibits any CO emission? If this were the case, star formation
in TDGs would have had to enrich the gas to its present 
metallicity.  We think that this scenario is unlikely because firstly,
it does not explain the common metallicity of TDGs. 
Secondly, star formation is occurring in many dwarf galaxies of 
sizes comparable
to TDGs (e.g. Taylor et al. 1998) at rates at least as 
high as in our TDG sample. 
Were star formation capable of enriching gas 
considerably, dwarf galaxies would not have such low metallicities.
Finally, it is hard to see
how molecular gas supposedly coming 
from the parent galaxies 
can survive the star formation necessary to increase its metallicity.


Thus we conclude that if large quantities of molecular gas had been
present in the outer parts of spiral galaxies -- from where the material
forming TDGs comes from -- it should be observable. This fact together
with the spatial and dynamical coincidence of atomic and molecular gas
suggests that we are seeing molecular gas formation.
The CO
emission furthermore coincides with the position of H$\alpha$ emission
tracing star formation. Thus, we are  witnessing the entire 
process of star formation: The formation of molecular from atomic gas
and the subsequent star formation.

\begin{figure}[t]
\label{mol_fraction}
\plotfiddle{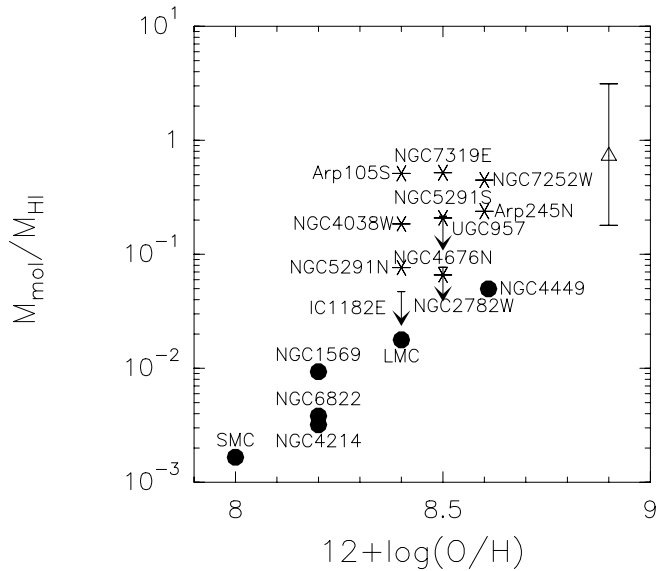}{7.5cm}{0}{50}{50}{-180}{-20}
\label{mol_fraction}
\caption{The molecular gas fraction versus metallicity.
For TDGs with no measured metallicity (see Tab. 1) we
have tentatively assumed 12+log(O/H) = 8.5.
The stars denote TDGs, filled 
circles classical dwarfs and the triangle the median value of the 
spiral  galaxies from the sample of Kennicutt (1998), the errorbar
showing the standard deviation around the median of that sample.}
\label{mol_fraction}
\end{figure}

\section{Star formation in TDGs, classical dwarfs and spirals}

Is star formation proceeding in the same way in TDGs as in classical
dwarfs and in spiral galaxies? In order to discuss this question we
compare some properties of our TDG sample to samples of classical
dwarfs and spirals. 
The data for spirals are taken from Kennicutt (1998). We calculated
median values and dispersions from his Table 1 and assumed
a metallicity of 12+log(O/H) =8.9 for this average spiral.
Details of the data of the classical dwarf sample 
are given in Braine et al. (2001), except for the 
HI data which is taken from Stil \& Israel (priv. communication, NGC~1569),
Melisse \& Israel (1994, NGC 4214), Israel et al. (1996, NGC 6822),
Hunter \& Thronson (1996, NGC 4449), Cohen et al. (1988, LMC) and 
Rubio et al. (1991, SMC). The HI masses refer to the area in which molecular
gas was detected.
The molecular gas mass for all objects was
calculated or scaled to the same conversion factor given in Table 1.

In Fig.\ 1
the molecular gas fraction versus metallicity is shown.
The molecular gas fractions of
TDGs are higher than for classical dwarfs and lie within 
or only slightly below the typical range of 
spiral galaxies. 
There seems to be a trend of \mmol/\mhi\, with metallicity, although
with a large scatter.
This trend most likely reflects a decrease of the 
CO-molecular-gas-mass conversion factor with decreasing metallicity
as indicated by
observational  (e.g. Wilson 1995) and  
theoretical (Maloney \& Black 1988) studies.

\begin{figure}[t]
\plotfiddle{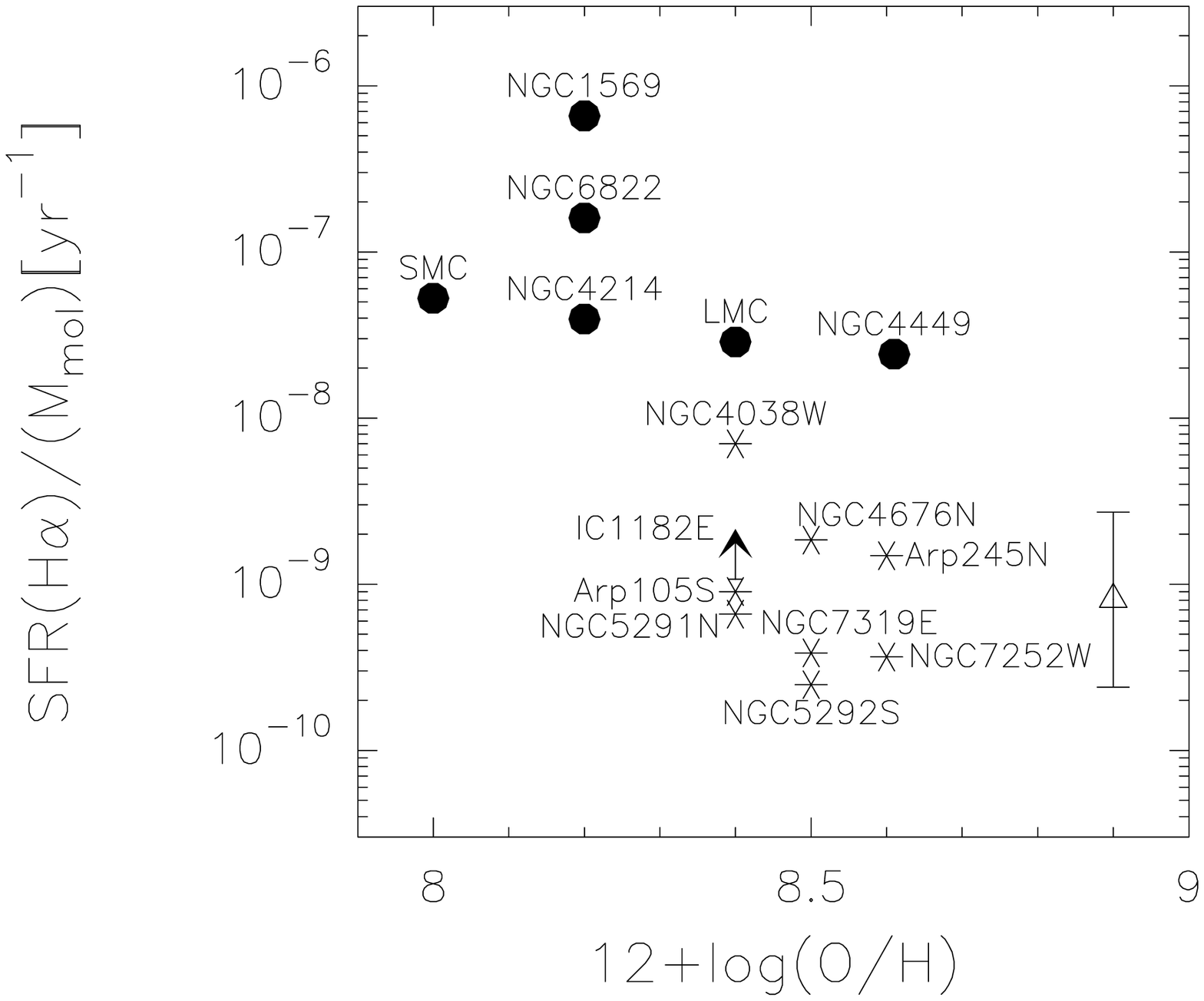}{2.5cm}{0}{40}{38}{-215}{-100}
\plotfiddle{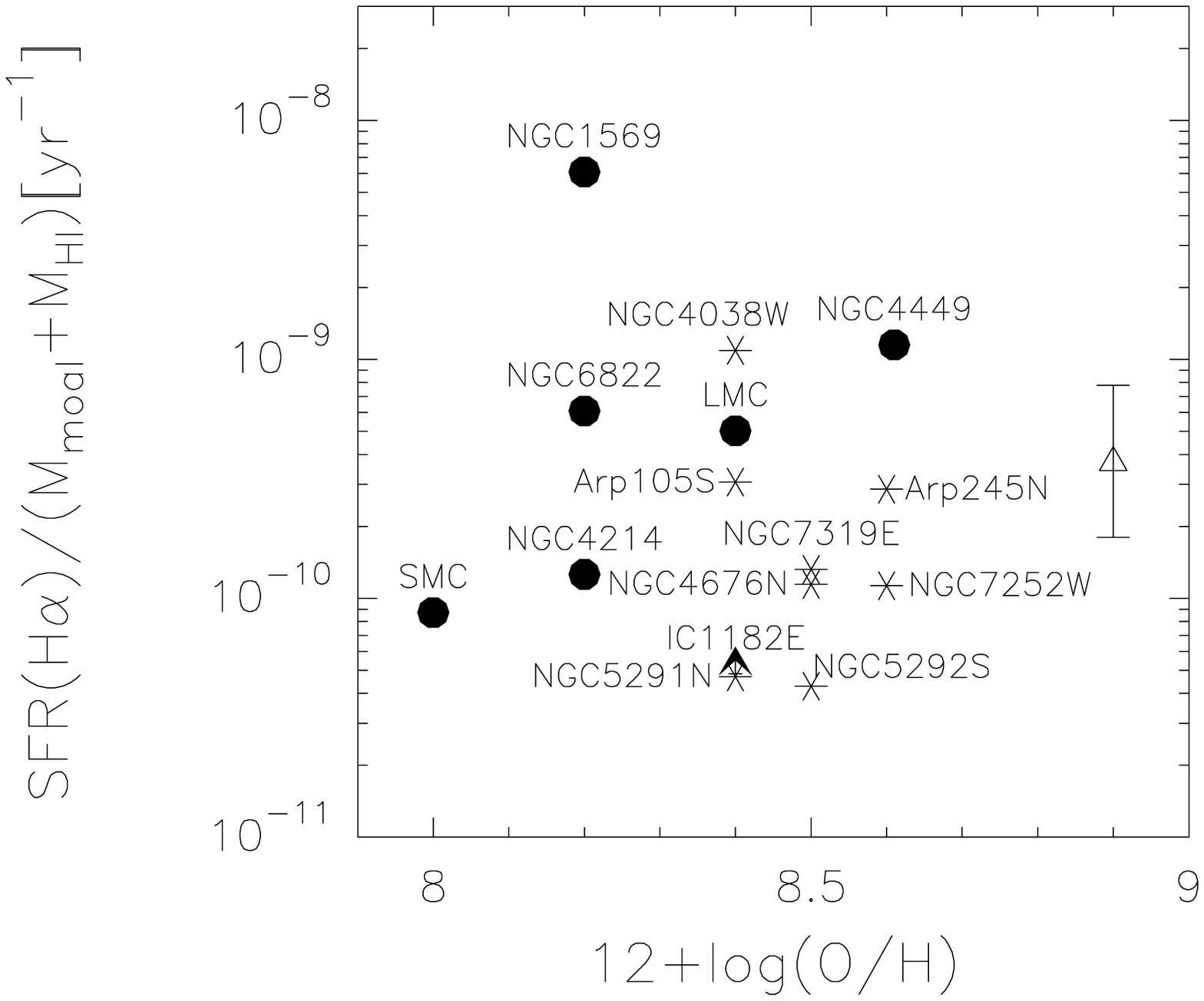}{2.5cm}{0}{40}{38}{-5}{-12}
\caption{{\bf Left}: The  SFR, traced by H$\alpha$ emission,
devided by the molecular gas mass. {\bf Right:} The SFR devided by the total
(molecular and atomic) gas mass. The different symbols describe the
three samples as explained in Fig.\ 1.}
\label{sfe}
\end{figure}

This means that  the molecular gas in TDGs is easier to detect than in
classical dwarfs due to the higher metallicity of TDGs.
Whereas classical dwarfs follow a luminosity-metallicity relation,
the metallicities of TDGs lie all in a narrow range between
8.4 and 8.6 (Duc et al. 2000). As a  consequence we are able
to observe 
molecular gas in TDGs of lower luminosities than classical dwarfs. 
In the latter objects no
CO is detected for metallicities lower than
12+log(O/H) = 7.9 (Taylor et al. 1998),  corresponding to galaxies
fainter than $M_B \approx -15$ mag (Skillman et al. 1989). 
TDGs of comparable luminosity can be observed in CO, as the detections
of NGC 7252W, NGC4038S and NGC7319E, all with blue magnitudes between
-14 and -15, show. 

\begin{figure}
\label{schmidt}
\plotfiddle{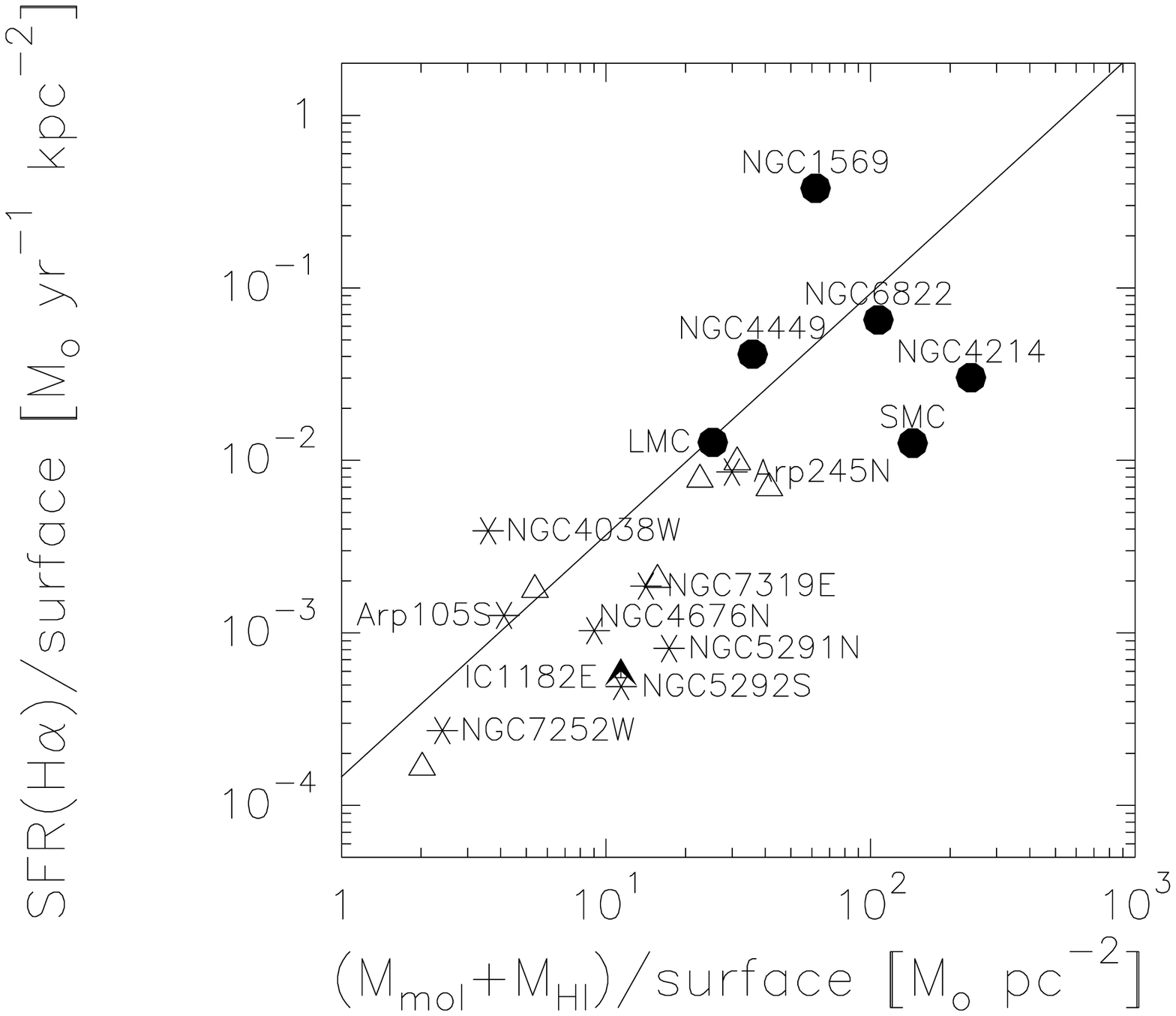}{7.cm}{0}{50}{50}{-180}{-20}
\label{schmidt}
\caption{The SFR per surface area  as a function of the gas mass surface
density. The solid line gives the best fit derived
by Kennicutt (1998) for a large sample of galaxies of different types (ranging
from dwarfs to starburst galaxies). The triangles are some galaxies
from Kennicutt's sample. }
\label{schmidt}
\end{figure}

A widely used parameter to study star formation 
is the star formation efficiency (SFE)
which is defined as SFR per molecular mass. Here, we use the H$\alpha$
luminosity as a tracer of the SFR as:
$
{\rm SFR} = 5 \times 10^{-8} (L_{H\alpha}/L_{\sun}) M_{\sun} {\rm} 
{\rm yr}^{-1}
$ 
(Hunter \& Gallagher 1986).
In Fig.\ 2 (left) we show the SFE versus metallicity.
There is 
a clear difference between TDGs and classical dwarfs:
Whereas the SFE of most TDGs falls  in the same range as for spirals, 
classical dwarfs possess a higher ratio. 
This difference persists even when considering a possible underestimate
of up to 50\% of the $L_{H\alpha}$ in some TDGs for which only long-slit
measurements are available.
Most likely the apparently high SFE of classical dwarfs is caused by an 
underestimate of the molecular gas due 
to their low metallicities. This assumption is supported
by Fig.\ 2 (right) where we plot the SFR devided
by the {\it total} gas mass. 
In this case the whole sample falls in a much narrower range.
The SFR normalized  to the {\it total} gas mass of some TDGs is 
somewhat lower than the range of values of the
spiral galaxy sample. This could be due to an 
overestimate of their HI content 
due to insufficient resolution of the data.
Alternatively, it could 
imply that TDGs  possess a reservoir of atomic gas that is not used
(yet) for star formation,  reflecting the fact that TDGs are in the process of
formation and the star formation is just starting.

Schmidt (1959) has suggested a relation between the SFR per surface area,
$\Sigma_{\rm SFR}$ and the gas mass
surface density, $\sigma_{\rm gas}$, $\Sigma_{\rm SFR} \propto 
 \sigma_{\rm gas}^\alpha$. 
Kennicutt (1998) showed that indeed a tight correlation between these two
quantities holds  for different types
of galaxies and spans many orders of magnitude.
In Fig.\ 3 we show this relation for our sample. We include the
best-fit line of the Kennicutt-sample. 
The entire  sample follow the same relation, indicating that star formation
proceeds in a normal fashion.

\end{document}